\documentclass{article}
\usepackage{graphicx}
\usepackage{amsmath,amssymb}
\usepackage{color}
\usepackage{natbib}
\usepackage{setspace}

\usepackage[hmargin=3cm,vmargin=3.5cm]{geometry}

\newcommand{\dy}[1]{\textcolor{black}{ #1}}
\renewcommand{\v}[1]{\ensuremath{\mathbf{#1}}} 
\title{Mixing and transport  by ciliary carpets: a numerical study}
\author{Yang Ding$^1$, Janna Nawroth$^2$, Margaret McFall-Ngai$^3$ \& Eva Kanso$^1$\thanks{Email address for correspondence: kanso@usc.edu} \\ $^1$Aerospace \& Mechanical Eng., University of Southern California, Los Angeles, CA 90089 \\ $^2$ Aerospace Laboratories, California Institute of Technology, Pasadena CA 91125\\ $^3$ Medical Microbiology \& Immunology, University of Wisconsin, Madison, Wi 53706
}

\begin{document}

	\maketitle
\abstract{
\singlespacing
We use a 3D computational model to study the fluid transport and mixing due to the beating of an infinite array of cilia. In accord with recent experiments, we observe two distinct regions: a fluid transport region above the cilia and a fluid mixing region below the cilia tip. The metachronal wave due to phase differences between neighboring cilia is known to enhance the fluid transport above the ciliary tip. In this work, we show that the metachronal wave also enhances the mixing rates in the sub-ciliary region, often simultaneously with the flow rate enhancement. Our results suggest that this simultaneous enhancement in transport and mixing is due to an enhancement in shear flow. As the flow above the cilia increases, shear rate in the fluid increases and such shear enhances stretching, which is an essential ingredient for mixing. Estimates of the mixing time scale indicate that, compared to diffusion, the mixing due to the cilia beat may be significant and sometimes dominates chemical diffusion.}

\section{Introduction}

Cilia are micro-scale hair-like structures that cover many eukaryotic cells, from single-celled protozoa to mammalian epithelial surfaces.  Motile cilia function in both fluid transport across the cell surface as well as in the sensing of environmental cues~\citep[][]{bloodgood2010sensory}. The distribution of cilia in animal tissues is highly correlated with the animal's habitat~\citep[][]{ibanez2003beat}. In aquatic species, cilia commonly occur along external and internal epithelial surfaces, where they have a broad array of functions, from food capture to acquisition of microbial partners. A well-studied example is the harvesting of the bacterium \emph{Vibrio fischeri}, the luminous symbiont of the Hawaiian squid, during the onset of the model squid-vibrio symbiosis~\citep[][]{altura2013first,kremer2013initial}. These bacteria aggregate along ciliated surfaces of a nascent light organ, where they present signals critical for normal onset and development of the symbiotic relationship.  When animals invaded terrestrial habitats, cilia became restricted to internal epithelial surfaces to reduce water loss across mucociliary membranes, rendering them difficult subjects for direct study. In mammals, they serve a number of functions including mucus clearance
in the respiratory system, left-right asymmetry determination during embryonic development and transport of egg cells in Fallopian tubes (see \cite{satir2007overview} and the references therein). Also, as in squid-vibrio system, mammals have intimate interactions with bacteria along their ciliary surfaces. In addition to their role in biological systems, cilia also provide an attractive paradigm for fluid manipulation in microfluidic devices.  This potential is particularly compelling in light of the rapid developments in micro-scale manufacturing technologies and applications such as ``lab-on-a-chip'' devices for synthesis of chemicals \citep[e.g.,][]{alexeev2008designing,shields2010biomimetic}.

\begin{figure}
\begin{center}
\includegraphics[width=1\textwidth]{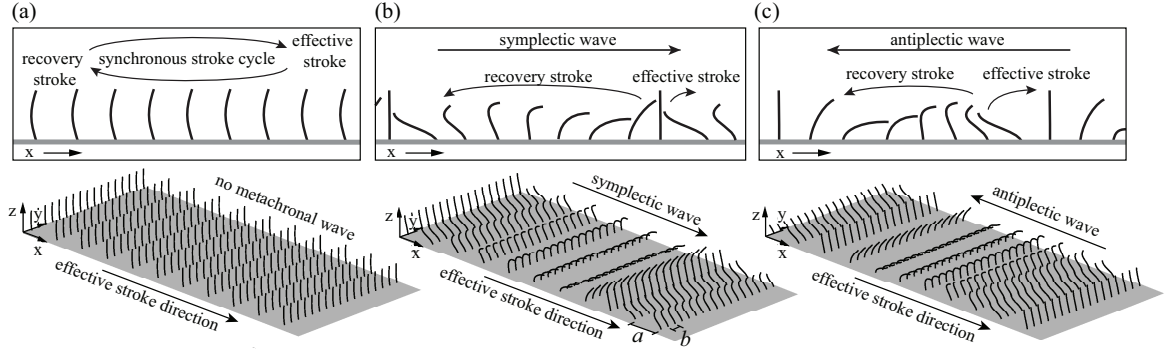}
\caption{{Schematic of ciliary array in (a) synchronized beating, (b) symplectic metachronal wave, and (c) antiplectic wave.  Top:  Single cilium or single row of cilia along x-axis. }Bottom: 2D ciliary carpet.
In the synchronized case, i.e. , with a phase difference of zero,
no metachronal wave is generated.  A negative phase difference creates symplectic waves due to compression of the cilia during the effective stroke phase. A positive phase difference results in antiplectic waves by compression on the cilia during the recovery stroke.
 \label{fig:cilia}}
\end{center}
\end{figure}

Cilia are usually $4$-$10$\,$\mu$m long in mammals and can reach up to $25$\,$\mu$m long in aquatic invertebrates such as the Hawaiian squid. Their slenderness ratio ranges between $10$ and $100$. Their beating motion usually consists of two asymmetric strokes: an effective stroke where the cilium is relatively straight and generates a flow in the same direction as its motion, and a recovery stroke where the cilium bends towards the surface and generates a relatively smaller backward flow \citep[see, e.g.][]{blake1974mechanics}. Cilia often beat in accord and form wave patterns, known as metachronal waves. A metachronal wave is called symplectic when it propagates in the direction of the effective stroke. It is called antiplectic when the wave propagates in the direction opposite to the effective stroke (Fig.~\ref{fig:cilia}). Cilia-driven flows are characterized by velocities less than $100$\,$\mu$m/s and oscillatory frequency less than $20$\,Hz~\citep[][]{blake1974mechanics}. The associated Reynolds numbers are typically about $10^{-3}$, even for fluid viscosities as low as water, so inertia is negligible and viscous forces are dominant.
Manipulating the surrounding fluid in this drag-dominated microscale work is not intuitive.
Cilia are successfully used by biological systems to manipulate the fluid at this scale and, as such,
have attracted a great deal of research effort in theoretical
and computational mechanics, \textcolor{black}{see, e.g., the reviews by \cite{blake1974mechanics}, \cite{brennen1977fluid}, and \cite{smith2008fluid}.  In particular, the coupling between the metachronal waves and the fluid motion was considered in a number of studies,  most of which focus on  either examining the role of the hydrodynamic coupling in generating the metachronal waves  
or examining the effect of the metachronal waves on the fluid transport.
In the following, we present a partial review of the related literature.}

\begin{figure}
\begin{center}
\includegraphics[width=0.3\textwidth]{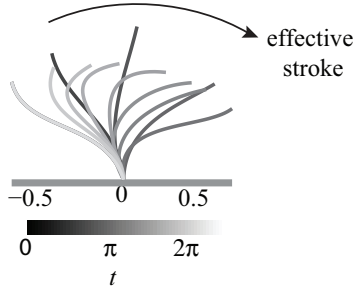}
\caption{Beating pattern of an individual cilium  based on \cite{fulford1986muco}. The effective stroke is shown in dark grey and recovery stroke in light grey.
 \label{fig:beatpattern}}
\end{center}
\end{figure}

\textcolor{black}{The `envelope' model where the tips of the cilia are modeled as a continuum was proposed by~\cite{blake1971spherical} and has been used in numerous studies
on the swimming of ciliated micro-organisms, see, e.g.,~\cite{michelin2010efficiency,michelin2011optimal}. The envelope model is only capable of describing symplectic metachronal waves and is only valid in the case of very densely spaced cilia.
Slender body theory, coupled with elastic models of the internal structure of the cilium, has been used to model individual and arrays of cilia.
For instance, it was used by \cite{gueron1992ciliary}, \cite{gueron1997cilia}, and \cite{gueron1999energetic} to demonstrate that, for closely packed ciliary arrays, hydrodynamic interactions between neighboring cilia are sufficient to account for antiplectic metachrony.
The role of hydrodynamic coupling in the emergence of the metachronal wave was also confirmed by~\cite{elgeti2013emergence}.~\cite{eloy2012kinematics} focused on computing
the energy-optimal kinematics of a wall-bound elastic cilium.~\cite{Osterman2011finding} used a measure of energetic efficiency based on the ciliary performance in pumping fluid and computed, according to this criterion, optimal
beating patterns of cilia acting individually and collectively.
The flow fields generated by beating cilia were examined numerically using the Stokeslet method \cite[see][and references therein]{smith2007discrete,ainley2008method,cortez2001method} as well as the immersed boundary method, see, e.g.~\cite{lukens:017511}
for a description of the flow field generated by a single cilium and~\cite{khaderi2011microfluidic,khaderi2012fluid} for the flow field generated by a finite number of artificial cilia.}


\textcolor{black}{These studies focus primarily on the performance of the ciliary systems in fluid transport and pumping.} However, the way by which ciliary motion serves the simultaneous tasks of fluid transport and sensing is less well understood. Transport implies that fluid is being moved above the
cilia tip whereas, for sensing, chemicals in the fluid need to be brought closer to the base of the ciliated surface.
It is unclear if chemical diffusion is the sole mechanism for sensing or whether cilia-generated flows enhance diffusion by chaotic advection and mixing. At this micro scale, lack of turbulence makes mixing by chaotic advection a nontrivial task \citep{aref1990chaotic,ottino1989kinematics}.
The computational results of \citet{lukens:017511} showed some evidence of mixing in the flow around a single cilium.
\textcolor{black}{A transition from unidirectional flow above the cilium to vortical flow below the cilium was reported by \cite{supatto2008all} based on in-vivo experiments that mapped the velocity field surrounding a single beating cilium in zebrafish embryo.}
Enhanced diffusion below the cilia tip was also reported in experiments on artificial cilia~\citep[e.g.,][]{shields2010biomimetic}, albeit for cilia beating in synchrony (no metachronal wave).
This suggests that the beating of ciliary carpets may be an effective way to induce chaotic mixing.

\textcolor{black}{The goal of this paper is, therefore, to obtain understanding of the transport and mixing properties of flows generated by a doubly-periodic array of asymmetrically beating cilia.
In particular, we ask how the transport and mixing properties of the ciliary-generated flows depend on the metachronal wave, and whether there is a trade-off between these two properties.}
To answer these questions, we use a computational model based on the regularized Stokeslet method (\S \ref{sec:model}).
We find, consistent with previous studies, that the flow is characterized by a transport region above the cilia tip and a mixing region below the
cilia tip with fluid ``leaking" between the two regions (\S\ref{sec:results}). We quantify the transport and mixing rates and their dependence on  the metachronal wave (\S\ref{sec:results}).
Comparing the mixing time scales to diffusion time scales, we conclude
that the mixing time scale is comparable to or
more favorable than that of pure diffusion (\S\ref{sec:conc}).

\begin{figure}
\begin{center}
\includegraphics[width=1.05\textwidth]{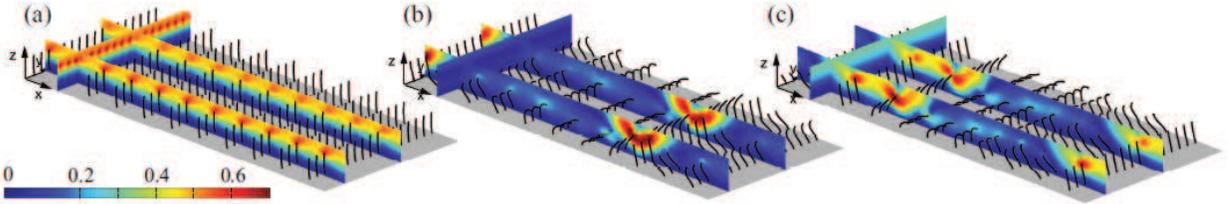}
\caption{Cilia array in (a) synchronized beating $\Delta \phi =0$, (b) symplectic metachronal wave with $\Delta \phi=-\pi/4$, and (c) antiplectic metachronal wave with $\Delta \phi=\pi/4$ (c). The color represents the magnitude of velocity field at three representative planes $y=0.08$, $y=3.00$, and $x=0$.  \label{fig:vfield3D}}
\end{center}
\end{figure}

\section{Model}
\label{sec:model}

We consider a regular array of cilia  whose base points are placed on the infinite plane
$z = 0$, with $(x,y,z)$ being the Cartesian coordinates. The spacing between the base
points of the cilia is $a$ in the $x$-direction and $b$ in the $y$-direction, see Fig.~\ref{fig:cilia}.
The length of each cilium is $\ell$.
 We assume that the cilia beat is in the $x$-direction with frequency $\omega$ and period $T=2\pi/\omega$,
 and that the metachronal wave is propagating in the
$x$-direction such that all cilia having the same value of $x$ are in phase with each other.
In order to model the metachronal wave, each cilium goes through exactly the same cyclic
motion as its neighbor but with a phase difference $\Delta \phi$.

For concreteness, consider the cilium whose base point is located at the origin. Let $\mathbf{x}^c \equiv (x^c,y^c,z^c)$
denote the coordinates of the centreline of this cilium.
The motion (effective and recovery stroke) of the cilium is described by $\mathbf{x}^c(s,t)$
where $s$ is the arclength along the cilium's centreline from its base $(0<s<\ell)$ and $t$ is time $( 0<t< T)$.
Throughout this work, we prescribe 2D kinematics ($y^c=0$) of the cilium using a Fourier
series expansion in $s$ and Taylor series in $t$ with coefficients based on experimental data, as shown in Fig.~\ref{fig:beatpattern},
(see \cite{fulford1986muco} for details). \textcolor{black}{This 2D beating kinematics is chosen to facilitate the analysis and the visualization of the flow field.
Previous studies suggest that, while 3D beating kinematics may be more efficient in transporting fluid, the mechanisms of generating a directional flow are the same in both 2D and 3D~\citep{elgeti2013emergence,eloy2012kinematics}. Therefore, we expect our results to be valid for 3D kinematics as well.}
The metachronal wave in the $x$-direction 
implies that  the kinematics of a cilium based at the point $(ma, nb, 0 )$, where $m,n \in \mathbb{Z}$, at time $t$ is given by
\begin{equation}
\mathbf{x}^c_{m,n}(s,t) = \bigl(ma+ x^c(s, {t}_m), \, nb, z^c(s, {t}_m)\bigr), \qquad
t_m =  \omega t + m \Delta \phi.
\end{equation}
The metachronal wave has a wavelength of $2\pi a /\Delta \phi$, frequency of $\omega/2\pi$ and wave velocity of $a \omega/\Delta \phi$. 

We use the cilia length $\ell$ to scale all length variables and $1/\omega$ to scale time. That is to say, all
variables are considered to be dimensionless with  $\ell =1$, $\omega =1$ and $T=2\pi$. To explore the effect
of the phase difference $\Delta \phi$ from 0 to $2\pi$ and avoid overlapping of neighboring cilia, we use a spacing $a=1.44$ in the $x$-direction and $b= 0.4$ in the $y$-direction, see Fig.~\ref{fig:cilia}b,\,c,\,\&\,d.

\begin{figure}
\begin{center}
\includegraphics[width=0.85\textwidth]{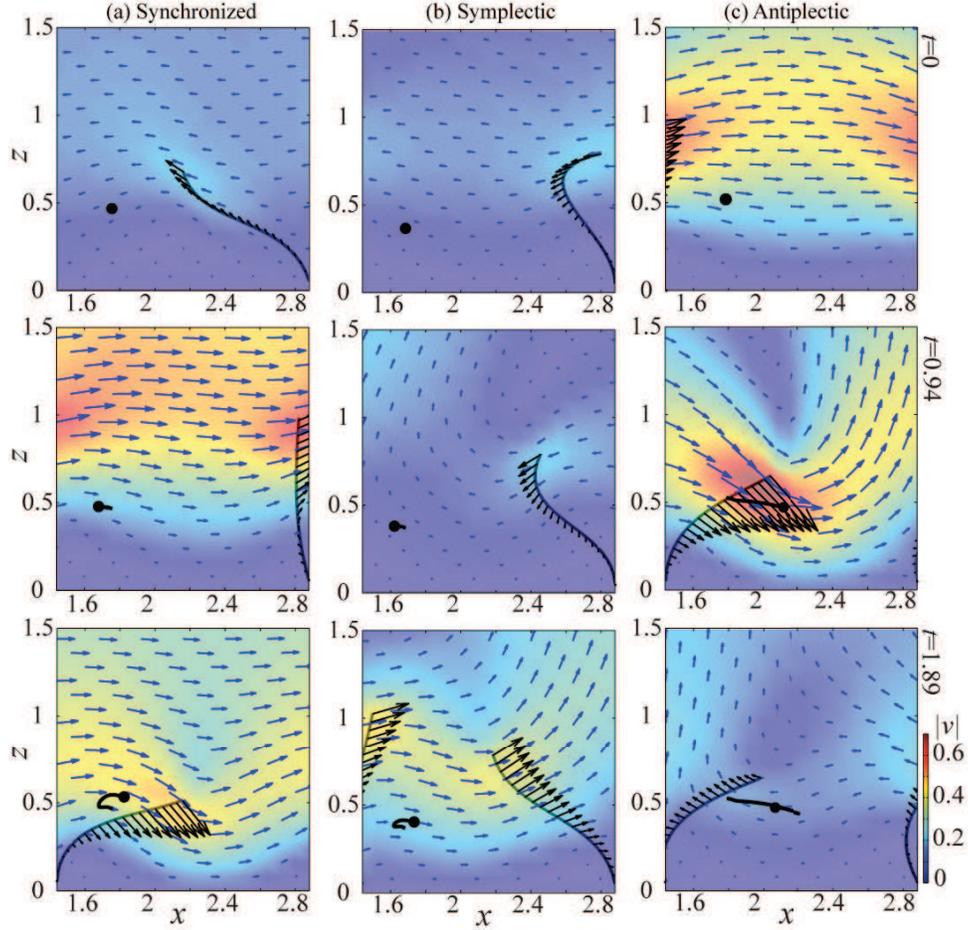}
\caption{Velocity field for (a) $\Delta \phi =0$, (b)  $\Delta \phi=-\pi/4$, and (c)  $\Delta \phi=\pi/4$
at three snapshots in time $t=0$, $0.94$, \& $1.89$, from top to bottom.
{The difference in the cilia configuration is due to the fact the window we show is not centered around the cilium at the origin. Except for the cilium at the origin, when beating in metachrony (with a phase lag), each cilium is at a different stage of the stroke cycle at the same time $t$.} Background color represents flow speed. The black circles represent tracer particles and their trajectories. {The black arrows represent the velocity of the cilia.}
\label{fig:vfield2D}}
\end{center}
\end{figure}

\begin{figure}
\begin{center}
\includegraphics[width=0.9\textwidth]{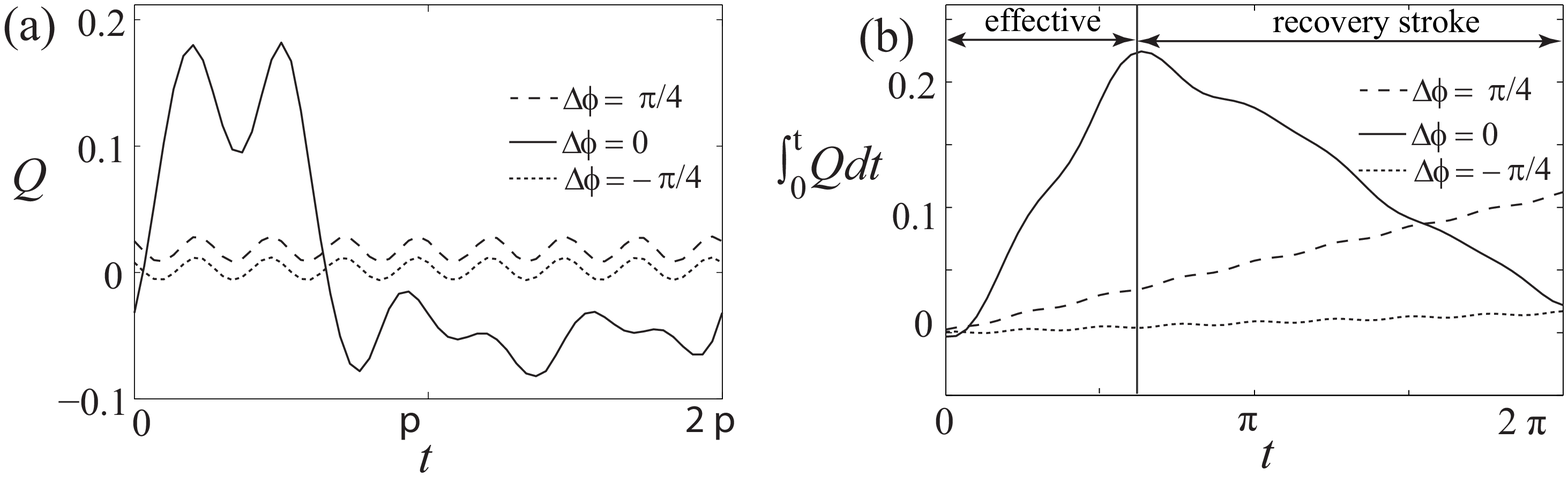}
\caption{Flow rate (a) and accumulated flow volume (b) as a function of time generated by synchronized beating, symplectic wave with $\Delta \phi=-\pi/4$, and antiplectic  wave with $\Delta \phi=\pi/4$.
\label{fig:flowvstime}}
\end{center}
\end{figure}
To compute the flow field generated by the ciliary motion, each cilium is approximated by a distribution of regularized Stokeslets
along its centreline, together with an ``image'' distribution to satisfy zero-flow boundary conditions at a plane wall~\citep{cortez2001method}.
{Note that the regularization parameter determines the size of a region with nearly uniform velocity, and therefore can be used as an approximation of the finite radius of the cilium~\citep{ainley2008method}. Therefore, we set the regularization parameter to $0.05$, which gives a radius-to-length ratio close to that of a typical cilium~\citep{blake1972model}.} 
 The Stokeslet strengths $\mathbf{F}$ depend on their position on the cilium and time and are obtained by imposing
no-slip boundary conditions at the cilia. Let $\mathbf{G}(\mathbf{x},\mathbf{x}_o)$
be the Green's  function associated with a regularized Stokeslet of unit strength situated at a point $\mathbf{x}_o$
and such that the no-slip condition is satisfied on the
plane $z = 0$. The total velocity induced by all cilia is given by
\begin{equation}
\mathbf{u}(\mathbf{x},t) = \sum_{n=-\infty}^{\infty}  \sum_{m=-\infty}^{\infty} \int_0^\ell
 \mathbf{G}(\mathbf{x}, \mathbf{x}^c_{m,n}(s,t)) \cdot \mathbf{F}(\mathbf{x}^c_{m,n}(s,t)) \mathrm{d}\,s,
 \label{u_F}
\end{equation}
An expression for $\mathbf{G}(\mathbf{x}, \mathbf{x}_o)$ is given by \citep{ainley2008method} \footnote{Note the expression (Eqn. 21) in \citep{ainley2008method}  has a sign error.}. No-slip boundary conditions imply that the fluid velocity  at the cilia centerline
is equal to the prescribed velocity of the cilia $\mathbf{u}(\mathbf{x}^c_{m,n},t) = {\mathrm{d}\mathbf{x}^c_{m,n}}/{\mathrm{d}\,t}.$
Substituting~\eqref{u_F} into this equation, one gets a linear system of equations to be solved for the Stokeslet strengths $\mathbf{F}$. Having
the Stokeslet strengths, the velocity field can be reconstructed everywhere.
{For more details on the regularized Stokeslet method, the interested reader is referred to~\cite{cortez2001method,ainley2008method}.}

{Given that we apply the regularized Stokeslet method to an infinite array of cilia,} a comment on convergence is in order here. Because of the presence of the plane wall, the velocity generated by a Stokeslet and its image decays faster than $1/r^2$, where $r$ is the distance from the Stokeslet. Therefore, the infinite summations in~\eqref{u_F} converge.
This is in contrast to an infinite summation over Stokeslets (no image system), which diverges because the flow given by a Stokeslet decays as $1/r$.
Techniques to overcome this challenge using Fourier transform and Ewald's summation date back to the work of \citet{hasimoto1959periodic} for point Stokeslets and have been recently extended to the context of regularized Stokeslets by \citet{leiderman2013regularization}.
However, since in the case considered here the infinite sums converge, we simply use truncated sums  $\sum_{n=-N_{\mathrm{max}}}^{N_{\mathrm{max}}}$ and $\sum_{m=-M_{\mathrm{max}}}^{M_{\mathrm{max}}}$ as approximations
of the infinite summations over $m$ and $n$. In other words, we approximate the infinite ciliary array
using a computational domain with a large but finite number of cilia.
To avoid any edge effects, we report the results for flow transport and mixing in the middle section of the computational domain.

\section{Results}
\label{sec:results}

The flow field generated by ciliary motion is studied for three distinct cases: (a) all the cilia beat together in a synchronized way; (b) the ciliary beat generates a symplectic metachronal wave that propagates in the same direction as the effective stroke of the individual cilium ($\Delta \phi$ is negative); (c) the ciliary beat generates an antiplectic wave propagating in the opposite direction of the effective stroke ($\Delta \phi$  is positive).
For each case, the flow field is depicted in Fig.~\ref{fig:vfield3D} at three representative planes $y = 0.08$, $y = 3.00$, and $x = 0$. Notice that
because the cilia are closely-spaced and their motion is synchronized in the $y$-direction, variations in the flow field in the $y$-direction are small and the flow field is nearly 2-dimensional.

A close-up on the velocity field near one of the cilia is depicted in Fig.~\ref{fig:vfield2D} at three distinct instances of the beat cycle (top to bottom).
The behavior of a passive tracer particle is superimposed.
{In the synchronized case, the instantaneous velocity field is in the same direction as the cilia motion. The high fluid velocities achieved during the effective stroke get reversed during the recovery stroke so that the net fluid flow over one cycle is small. In the two cases of the metachronal beating, the instantaneous velocity field is attenuated but, at the same time, the variation in the instantaneous velocities is smaller. This results in larger net flow over one cycle in the antiplectic case. Roughly speaking, in the antiplectic case, the cilia undergoing recovery stroke `cluster' together, thus minimizing reverse flow whereas in the  symplectic case, the cilia undergoing effective stroke cluster together thus minimizing flow in the effective stroke direction. To quantify these statements, we compute the volumetric flow rate generated by the cilia as a function of time. To this end, we adapt a formula for the flow rate generated by a point force near an infinite wall given by \citet{smith2008fluid}, namely, $Q=\frac{1}{\mu\pi}F_x z$, where $\mu$ is the viscous coefficient, $F_x$ is the force in the flow direction, and $z$ is the distance of the point force from the wall. The flow rate generated by the cilia in Fig.~\ref{fig:vfield2D} is plotted in Fig.~\ref{fig:flowvstime}(a) as a function of time over one ciliary beat cycle. Clearly, during the effective stroke, the instantaneous flow rate is higher for the synchronized cilia than the metachronally-beating cilia. However,  the flow rate is lower during the recovery stroke such that the net flow rate $\int_0^T Q dt$ over one ciliary beat cycle is small, see~\ref{fig:flowvstime}(b). }
Similar results were obtained by~\cite{khaderi2011microfluidic,khaderi2012fluid}, who noted that the blocking effect of reverse flow by the antiplectic metachronal wave is the mechanism the cilia use to enhance fluid transport.

\begin{figure}
\begin{center}
\includegraphics[width=0.9\textwidth]{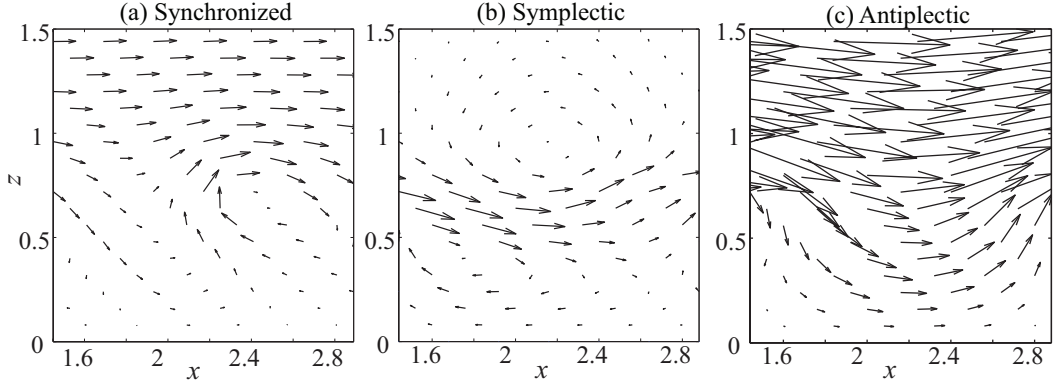}
\caption{Displacement field in a vertical plane ($y=0.08$) generated by (a) synchronized beating; (b) symplectic wave with $\Delta \phi=-\pi/4$; and (c) antiplectic  wave with $\Delta \phi=\pi/4$. \label{fig:poincare}}
\end{center}
\end{figure}

\begin{figure}
\begin{center}
\includegraphics[height=0.33\textwidth]{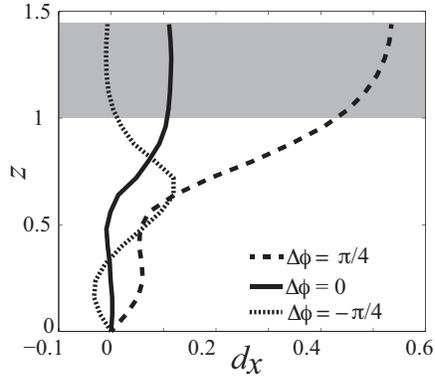}
\caption{Average displacement in the $x$-direction as a function of height for: synchronized $\Delta \phi = 0$, symplectic $\Delta \phi = -\frac{\pi}{4}$ and antiplectic $\Delta \phi = \frac{\pi}{4}$.
\label{fig:vprofile}}
\end{center}
\end{figure}

\begin{figure}
\begin{center}
\includegraphics[width=0.9\textwidth]{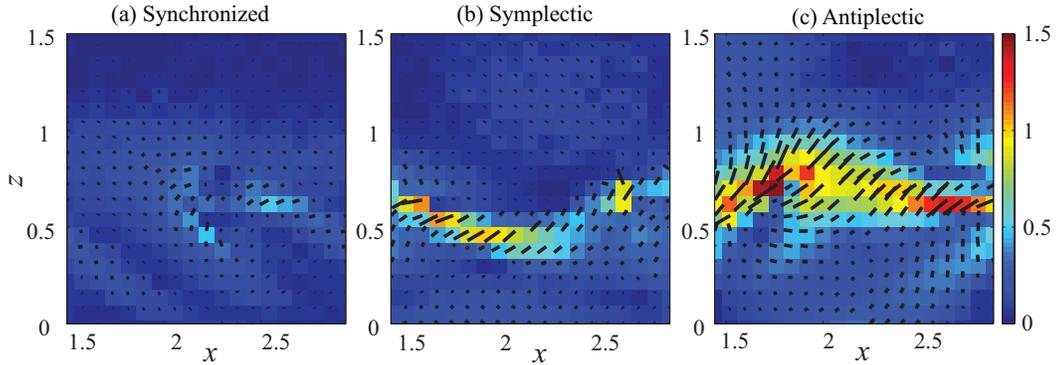}
\caption{Stretching of the fluid in the vertical plane for (a) synchronized beating $\Delta \phi =0$, (b) symplectic wave with $\Delta \phi=-\pi/4$, and (c) antiplectic wave with $\Delta \phi=\pi/4$. The color map represents the stretching magnitude whereas the orientation of the black lines indicates the direction of the stretching.
\label{fig:stretch_rates}}
\end{center}
\end{figure}

The displacement of a tracer particle over a cycle is given by $\mathbf{d}(\mathbf{x})=\int_0^T \mathbf{u}(\mathbf{x}(t),t) \mathrm{d}t$.  The displacement field is depicted in Fig.~\ref{fig:poincare} for the three cases shown in Fig.~\ref{fig:vfield2D}.
Above the cilia tip, tracer particles move in the $x$-direction almost uniformly in the synchronized and antiplectic cases with larger magnitude in the antiplectic case. Below the cilia, in the synchronized case, the displacement field shows a vortex-like structure whereas in the antiplectic case, there is little or no backward displacement. In the symplectic case, vortex-like structures appear both below and above the cilia tip with a narrow band of forward displacement in between. Note that the details of these vortex-like structures depend on $\Delta\phi$.
For example, for  $\Delta \phi < - \pi/3$, the vortex above the cilia tip disappears and the displacement field is analogous  to that of the synchronized case.


We compute the average displacement in the $x$-direction as a function of $z$ using \textcolor{black}{$\langle d_x \rangle(z)=\frac{1}{\lambda b} \int_0^{\lambda} \int_0^{b} d_x \mathrm{d}x \mathrm{dy}$ where $d_x$ is the $x$-component of the displacement field $\mathbf{d}$ and $\lambda$ is the wavelength of the metachronal wave}. The dependence of $\langle d_x \rangle$ on $z$ is depicted in Fig.~\ref{fig:vprofile}(a).
The average displacement is small close to the surface due to the no-slip boundary condition. It reaches a plateau above the cilia tip, and decreases for large $z$ further away from the cilia (results not shown).
Such a displacement profile is reminiscent to that of a shear flow. Clearly, the details from $z=0$ to $z\approx 1.2$ (at which the displacement plateaus) depend on $\Delta \phi$.

\begin{figure}
\begin{center}
\includegraphics[width=1\textwidth]{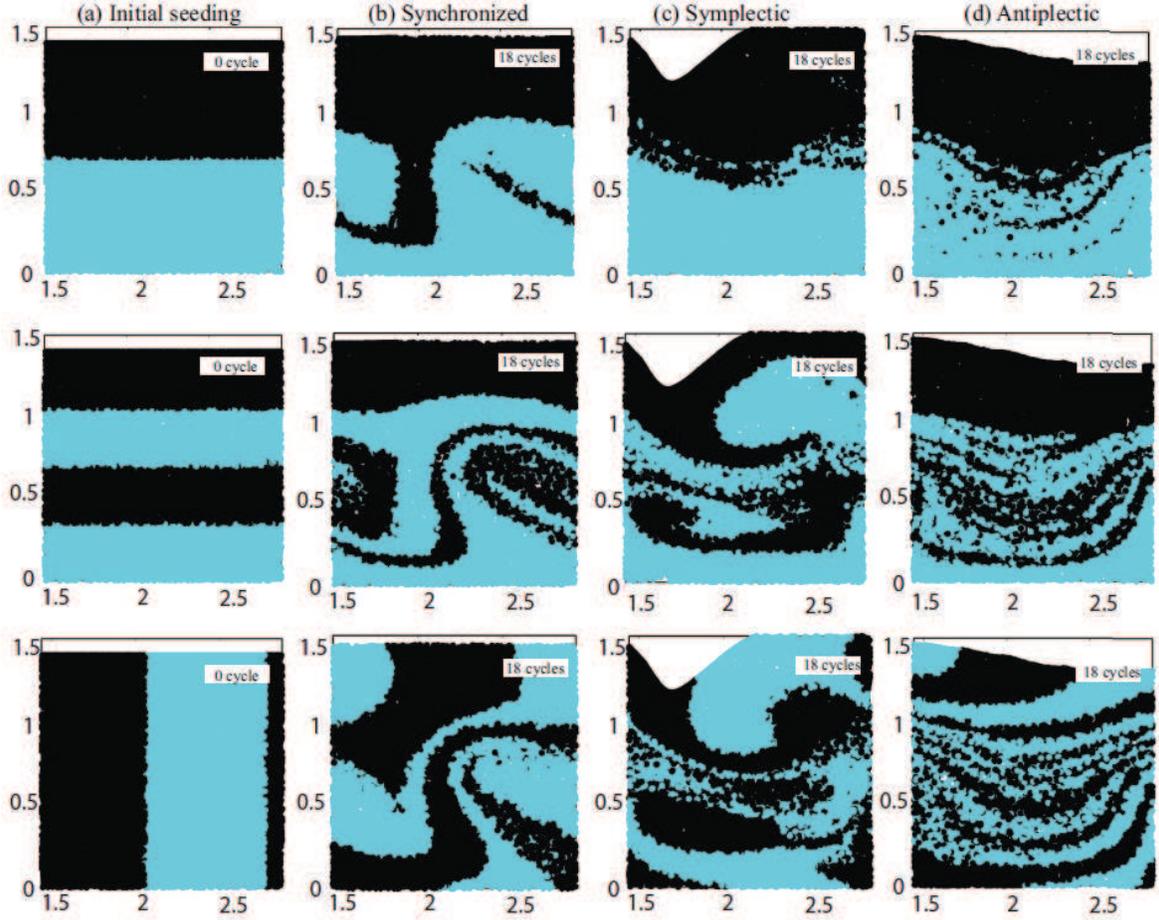}
\caption{Poincar\'{e} sections  for (a) three different initial seedings for the three representative cases: (b) $\Delta \phi=0$, (c)  $\Delta \phi=-\pi/4$, and (d) $\Delta \phi=\pi/4$.
\label{fig:mixsnapshot_vhalf}}
\end{center}
\end{figure}

\begin{figure}
\begin{center}
\includegraphics[width=1\textwidth]{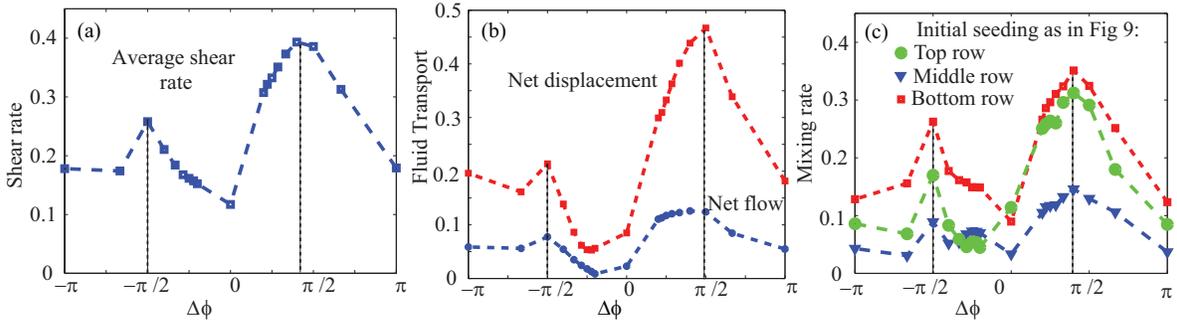}
\caption{(a) Average stretching rate over the entire domain as a function of the phase difference $\Delta \phi$ between adjacent cilia. (b) Net forward displacement and net volume flow generated per cycle as a function of the phase difference $\Delta \phi$.
(c) Mixing rate as a function of the phase difference $\Delta \phi$.
All three exhibit local maxima near $\Delta \phi=\pi/2$ (antiplectic wave) and  near $\Delta \phi=-\pi/2$ (symplectic wave).
\label{fig:mixing_rates}}
\end{center}
\end{figure}

{In order to examine the shearing properties of the displacement field shown in Fig.~\ref{fig:poincare},
we compute the gradient of the displacement field $\nabla \mathbf{d}$. We average the $y$-component of the displacement field $\mathbf{d}$ and consider only the gradient in the $(x,z)$ plane. The gradient $\nabla \mathbf{d}$ can be decomposed into two components: an antisymmetric component $\mathbf{R} = (\nabla \mathbf{d} - (\nabla \mathbf{d})^T)/2$, where $()^T$ is the transpose,  that corresponds to rotations, and a symmetric component, $\mathbf{S}=(\nabla \mathbf{d} + (\nabla \mathbf{d})^T)/2$, that corresponds to shear deformations. The two eigenvalues of $\mathbf{S}$ are of the form $\pm \lambda$ (by incompressibility) and indicate the rates of stretching ($+\lambda$) and compression ($-\lambda$). The unit eigenvector $\mathbf{e}_\lambda$ corresponding to the positive eigenvalue $(+\lambda)$ indicates the direction of stretching. Note that the eigenvector corresponding to $-\lambda$ is orthogonal to $\mathbf{e}_\lambda$, hence it suffices to show the vector field associated with $\lambda\mathbf{e}_\lambda$.  In order to understand the significance of the vector field $\lambda \mathbf{e}_\lambda$, recall that in the case of a uniform linear shear profile  ${d}_x = c z$ (where $c$ is a constant), $\lambda \mathbf{e}_\lambda$ is a uniform vector field of magnitude $c/2$ oriented at a $45^\circ$ to the $x$-direction.  The stretching rate and direction $\lambda\mathbf{e}_\lambda$ are depicted in Fig.~\ref{fig:stretch_rates} for the three cases shown in Fig.~\ref{fig:poincare}. In all three cases, the stretching rate is highest near the upper part of the cilia, and the stretching rate is enhanced by the metachronal wave, especially in the antiplectic case $\Delta \phi = \pi/4$. The $\lambda \mathbf{e}_\lambda$ field is oriented at a $45^\circ$ to the $x$-direction in regions where the shear rate $\lambda$ is maximum, but, overall, the stretching orientation is a nonlinear function of space. The change in the stretching orientation as a function of $(x,z)$ indicates the presence of {\em `folding'} in the displacement field $\mathbf{d}$ \citep{kelley2011separating}. We will return to this point later.
In Fig.~\ref{fig:mixing_rates}(a), we show the spatially-averaged shear rate over the  $(x,z)$ domain as a function of the phase lag. The average shear rate is characterized by two local maxima, one corresponding to a symplectic wave ($\Delta \phi \approx -\pi/2$) and another one corresponding to an antiplectic wave ($\Delta \phi \approx \pi/2$).}


We now return to our discussion on fluid transport. To assess the magnitude of the fluid transport in the $x$-direction, we use two interdependent quantities:
(i) the net displacement above the cilia tip and (ii) the net volume flow. We define the net displacement
 as $\frac{1}{0.4}\int_1^{1.4} \langle d_x \rangle dz$, which corresponds to the net displacement in the grey region of Fig.~\ref{fig:vprofile}.
We calculate the time-averaged net flux generated per cilium as $\frac{1}{T}\int_0^T Q \mathrm{d}t$, where $Q$ is adapted from~\citep[][]{smith2008fluid} as noted earlier.
The dependence of fluid transport in the $x$-direction on $\Delta \phi$  is depicted in Fig.~\ref{fig:mixing_rates}(b). Both quantities, the net  displacement and net flow, exhibit a similar dependence on $\Delta \phi$.  It is clear that
both symplectic and antiplectic waves can enhance fluid transport in comparison with the synchronized case
($\Delta\phi = 0$) but, for small phase lag, the symplectic wave does slightly reduce the fluid transport. Maximal transport occurs
for antiplectic waves near $\Delta \phi=\pi/2$ and a smaller peak occurs for symplectic waves near $\Delta \phi=-\pi/2$.

The displacement field $\mathbf{d}(\mathbf{x})$ can be used as a discrete map $\mathbf{x}(t+T) = \mathbf{x}(t) + \mathbf{d}(\mathbf{x}(t))$ to study the long-term behavior of tracer particles. The behavior of tracer particles under repeated iterations of this map is equivalent to a Poincar\'{e} section associated with the underlying flow field. Poincar\'{e} sections of  the flow fields generated by the synchronized, symplectic and antiplectic ciliary beat are shown in  Fig.~\ref{fig:mixsnapshot_vhalf} for three different initial seeding of particles.
Particles of two different colors are used such that initially, they occupy equal volumes, that is to say, the two colors are in 50:50 ratio.
In Fig.~\ref{fig:mixsnapshot_vhalf} (top row), the initial seeding is chosen such that the black particles occupy the region between $z=0.7$ and $z=1.4$. Note that $0.7$ is chosen because it corresponds to the average height of the cilium over a cycle  (the actual length of the cilium is $1$ but its average height is smaller due to its bending motion during the effective and recovery strokes).
Clearly, the tracer particles from the upper layer leak into the lower layer.
This is also evident in Fig.~\ref{fig:mixsnapshot_vhalf}(middle row) with initial seeding consisting of four layers of alternating colors from  $z=0$ to $z=1.4$.
Again, black particles from the transport region between $z=1.05$ and $z=1.4$ leak and mix with subciliary particles. The
two regions, namely, the transport region above the cilia tip and mixing region below, are most distinguishable in Fig.~\ref{fig:mixsnapshot_vhalf}(bottom row) where the initial seeding consists of regions of alternating colors in the $x$-direction.  We emphasize that these two regions are not separated by clear boundaries and that fluid particles leak between them over the course of multiple stroke cycles.

\begin{figure}
\begin{center}
\includegraphics[width=\textwidth]{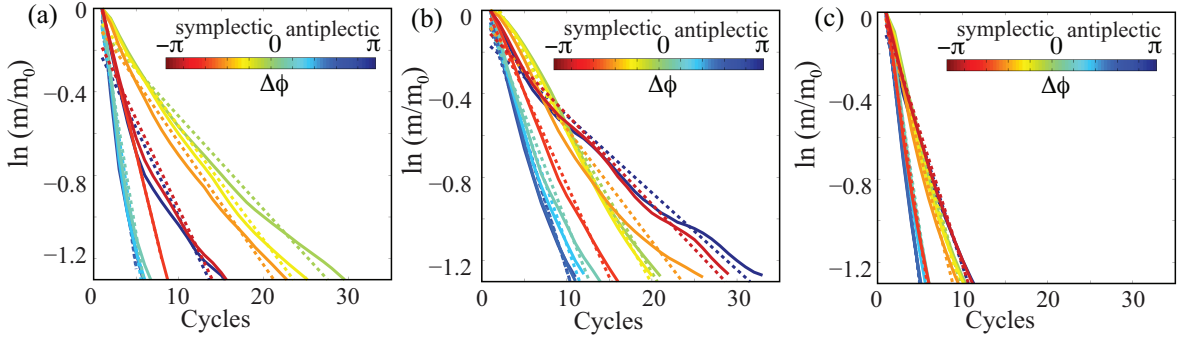}
\caption{Mixing rates:  Mixing numbers normalized to their initial value as a function of number of cycles for the three initial particles seeding shown in Fig.~\ref{fig:mixsnapshot_vhalf}.}
\label{fig:mixing_3cases}
\end{center}
\end{figure}

A common way to measure mixing is to quantify how the distance between these initially separated and differently-colored tracer particles decays as a function of time. Several (somewhat related) measures of mixing exist; see, for example, \cite{mathew2005multiscale,wiggins2004foundations} and references therein. Here, use the shortest distance between particles with different colors as a measure, following \cite{stone2005imaging}, and define the mixing number as $m=(\prod_{i=1}^{i=N}{ \mathrm{min}(|\v{x}_i-\v{x}_j|)^2 } )^{\frac{1}{N}}$, where $\v{x}_i$ and $\v{x}_j$ are positions of tracer particles of different colors and $j=1,2,...N$ ($N$ is the total number of same-color particles). Geometric mean is used here to better account the changes to the particles with small distances.
We find (see Fig.~\ref{fig:mixing_3cases}) that $m$ decreases with the number of cycles indicating enhanced mixing. {The decay in $m$ seems exponential, which is an indication of chaotic mixing. The mixing number $m$  is ill-defined for large times when the separation distance between particles of different colors gets closer to the spatial resolution of the initial seeding (which is $0.04$). For the time scales before this limit is reached, we approximate the decay by an exponential fit using $m/m_0=e^{-\eta_m N}$ with $m_0$ being the initial mixing number, $\eta_m$ the mixing rate, and $N$ the number of cycles.  
We use the resulting mixing rate $\eta_m$ to  compare the mixing efficiency as a function of the phase lag $\Delta \phi$.
The dependence of the mixing rate $\eta_m$ on $\Delta \phi$ is depicted in Fig.~\ref{fig:mixing_rates}.
Interestingly, we find that the values of $\Delta \phi$
that correspond to optimal transport (Fig.~\ref{fig:vprofile}(b)) are also optimal for mixing.}

{It is worth noting that, for certain phase differences $\Delta \phi$, the actual decay deviates slightly from the exponential fit, especially as $N$ increases (see Fig.~\ref{fig:mixing_3cases}). We attribute this decrease in the mixing rate to the no-slip boundary condition at the fixed wall, which is known to slow down the mixing rate, see, e.g., \cite{thiffeault2011moving} and references therein. \cite{thiffeault2011moving} propose moving the wall as a mechanism for attenuating the effect of the no-slip wall and restoring the exponential decay in the bulk of the fluid.  In many systems, it is not biologically viable to move the cell wall to which the cilia are attached. However, by design, the cilia during the recovery stroke align themselves almost tangentially to the fixed wall and, roughly speaking, act as a moving wall parallel to the no-slip fixed wall, thus shielding the fluid from the presence of the no-slip fixed wall and attenuating its effect on the fluid motion. The amount of `coverage' of the no-slip fixed wall by the moving cilia depends on the phase difference $\Delta \phi$, and therefore we observe that, for certain $\Delta \phi$, this attenuation is smaller and the mixing rate is not exactly exponential.
This slight deviation from the exponential decay does not affect our comparative study of the effect of the metachronal wave on the mixing rate, given that we are mainly interested in a relative measure of mixing as $\Delta \phi$ varies and not its absolute details.}

These results on the role of the metachronal wave in enhancing the fluid mixing  can be understood by recalling that stretching and folding are the building blocks of mixing.
The higher mixing rate associated with antiplectic waves is contributed to the higher shear rate  and therefore, higher stretching (see Figs.~\ref{fig:vprofile} and~\ref{fig:stretch_rates}). This shear rate argument also explains
why for $\Delta \phi$ between $-\pi/4$ and $0$, the fluid transport is reduced compared to the synchronized case but the mixing rate is enhanced (Fig.~\ref{fig:mixing_rates}). Indeed,
for these values of $\Delta \phi$, although the net flow is small, the rate of change in the displacement as a function of height (Fig.~\ref{fig:vprofile}) and consequently the average shear rate (Fig.~\ref{fig:mixing_rates}(a)) are greater than that in the synchronized case.

%
%

\begin{figure}
\begin{center}
\includegraphics[width=0.4\textwidth]{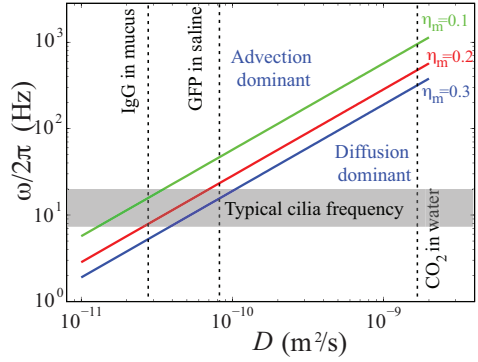}
\caption{Ciliary beat frequency versus diffusivity coefficient. The lines show when the mixing time scale due to diffusion is equal to that due to  advection by ciliary beating.  Representative mixing rates from Fig.~\ref{fig:mixing_rates}(c) are chosen such that $\eta_m$=0.1, 0.2 \& 0.3. The gray region indicates the typical beating frequencies of cilia.  Dashed lines indicate the diffusion coefficients for human IgG in mucus~\cite[][]{saltzman1994antibody}, green florescent protein (GFP) in aqueous saline~\cite[][]{swaminathan1997photobleaching}, and CO$_2$ in water~\cite[][]{fridlyand1996quantitative}.
 \label{fig:diffusion3}}
\end{center}
\end{figure}

{We conclude this section by comparing the time scales associated with mixing by fluid advection to those associated with mixing by molecular diffusion.
This comparison is motivated by the fact that  cilia often serve two biological functions: fluid transport
and sensing. For sensing, chemicals have to come close to the base of the ciliated surface,
where molecules would diffuse into the ciliated cells. In the absence of fluid motion, the only mechanism for sensing is molecular diffusion. The main question here is
whether the flows generated by the ciliary motion results in enhancing the time scales associated with sensing.
To this end, consider particles of different colors initially at a distance $s_o$ apart. We seek to compare the time scales for which the distance between them decreases by, say,
a ratio of $\alpha$, where $\alpha \in (0,1)$. From the standpoint of mixing by fluid advection, assuming it takes $N$ cycles for the separation distance between the particles to become $s_N = (1-\alpha) s_o$ and assuming exponential decay in particle separation distance $s^2_N = s^2_o e^{-\eta_m N}$, one gets that the mixing time scale
is given by}
\begin{equation}
{t_{\rm mixing} = \dfrac{2\pi N}{\omega} = -\dfrac{4\pi\log(1-\alpha)}{\eta_m \omega},}
\end{equation}
{where $\omega$ is the frequency of the ciliary beat, as defined in~\S\ref{sec:model}. On the other hand, particles moving a distance $\alpha \, s_o$
by molecular diffusion with diffusivity coefficient $D$ would take, on average,}
\begin{equation}
{t_{\rm diffusion} = \dfrac{\left( \alpha \, s_o\right)^2}{D}.}
\end{equation}
{Equating the two time scales $t_{\rm mixing} = t_{\rm diffusion}$, one gets}
\begin{equation}
{ \omega = \dfrac{4\pi \log(1-\alpha)}{ \left( \alpha \, s_o\right)^2 \eta_m} D.}
 \label{eq:advdiff}
\end{equation}
{For given $\alpha$, $s_o$ and $\eta_m$ (dictated by $\Delta \phi$), \eqref{eq:advdiff} provides a linear relationship between $\omega$ and $D$ which defines the boundary in the parameter space $(D,\omega)$ between advection-dominant mixing and diffusion-dominant mixing. This parameter space  is depicted in Fig.~\ref{fig:diffusion3}
for  $\alpha = 0.9$, $s_o=l=10\,\mu$m, which is the length of a typical cilium, and $\eta_m=0.1,\,0.2, \, 0.3$, which are representative values of the mixing rates obtained from Fig.~\ref{fig:mixing_rates}(c). One can readily see that there exists a range of diffusivity coefficients and ciliary beat frequencies where the time scales associated with mixing by fluid advection is dominant.}

%
%
%

\section{Conclusions}
\label{sec:conc}

We examined the fluid transport and mixing induced by ciliary carpets. Our results can be summarized as follows:
\begin{itemize}
\item[(i)] The fluid motion can be described by two distinct regions: a fluid transport region above the cilia and a fluid mixing region below the cilia tip, with fluid particles leaking between the two regions.
This is in accord with the experimental work on artificial cilia beating in synchrony~\citep{shields2010biomimetic}.
\item[(ii)] We showed that both transport and mixing can be enhanced, often simultaneously, when the cilia beat produces a metachronal wave. Further, the optimal wavelengths for transport and mixing are practically the same. Two optima are identified: one for antiplectic waves $\Delta \phi \approx \pi/2$ and one for symplectic waves $\Delta \phi \approx -\pi/2$, with better transport in the antiplectic case but comparable mixing in both cases.

\item[(iii)]
{Our fluid transport results are qualitatively consistent with the previous studies of \cite{gauger2009fluid} and \cite{khaderi2012fluid}. A direct numerical comparison is not feasible due to differences in the cilia beat stroke, density and shape. However, a rough comparison between these various studies shows some consistency in the resulting flow rates. In \cite{gauger2009fluid}, the normalized flow rate per cycle generated by the optimal beating pattern of a single cilium is close to $0.09$/cycle. In our model, we obtain a flow rate of $0.13$/cycle for a single cilium, whereas \cite{khaderi2012fluid} obtained a flow rate close to $0.1$/cycle per cilium for a finite number of ``flat" cilia  (cilia with plate-like geometry as opposed to circular cross-section). However, while \cite{khaderi2012fluid} report an enhancement in the flow rate up to 2 times due to the metachronal wave,  in our study, we observe a greater enhancement (up to 4 times) in the flow rate. We attribute this more significant enhancement in the flow rate to the higher cilia density in our study; other factors such as beating pattern may also contribute to the difference.}
\item[(iv)] \dy{To our knowledge, our study is the first that quantifies mixing by chaotic advection in a doubly-periodic array of cilia.
The possibility of mixing by chaotic advection below the cilia tips is suggested in the experimental studies of~\cite{supatto2008all} on a single beating cilium in zebrafish embryo and~\cite{shields2010biomimetic} on arrays of artificial cilia beating in synchrony.
More concrete evidence of mixing by chaotic advection was reported in the computational studies of \cite{lukens:017511} on a single cilium in a two-dimensional fluid and \cite{khatavkar2007active} on single and two cilia in a  two-dimensional rectangular channel.
In the case of two cilia (micro-actuators) beating at a  phase difference $\Delta \phi$, \cite{khatavkar2007active} reported exponential growth in the length of a fluid strip for $\Delta \phi$ between $\pi/3$ and $2\pi/3$ with the optimal value being $\pi/2$, which is consistent with our results.}
\item[(v)]
We explained the mixing below the cilia tip by the stretching and folding properties of the flow field generated by the ciliary array: the no-slip boundary condition at the base surface together with the asymmetric stroke of the beating cilia generate a shear-like flow field with stretching and folding as depicted in~\ref{fig:stretch_rates}. This is most likely a general phenomenon in all ciliary beat patterns.
\end{itemize}

We conclude by noting that, whereas all the results shown here were based on 2D beating kinematics, the computational framework that we use is general and can easily incorporate 3D ciliary motion. Indeed, our preliminary results on 3D nodal cilia show qualitatively similar separation of transport and mixing regions and similar effects due to the metachronal waves. We limited the discussion here to 2D beating kinematics to better illustrate the main ideas and for easier visualization of  the resulting flows.
Future extensions of this work will include coupling these models to advection-diffusion models for chemicals \citep[e.g.,][]{michelin2010efficiency}, as well as studying the transport and mixing of non-passive inanimate and motile particles in the presence of mucus (non-Newtonian fluids). 


\section{Acknowledgement}
We thank the Center for High-Performance Computing and Communications at USC for providing HPC resources and support.
\bibliographystyle{jfm}
\bibliography{./masterbib}

\end{document}